\documentclass[twocolumn,preprintnumbers,amsmath,amssymb,showpacs,aps]{revtex4}
\usepackage{graphicx}
\usepackage{dcolumn}
\usepackage{bm}
\usepackage{amsmath}
\usepackage{amssymb}
\usepackage{epsf}
\usepackage{epstopdf}
\newcommand{\be}{\begin{equation}}
\newcommand{\ee}{\end{equation}}
\newcommand{\bea}{\begin{eqnarray}}
\newcommand{\eea}{\end{eqnarray}}

\begin{document}
\title{Low-energy molecular collisions in a permanent magnetic trap}
\author{Brian C. Sawyer}
% \email{sawyerbc@colorado.edu}
\author{Benjamin K. Stuhl}
\author{Dajun Wang}
\author{Mark Yeo}
\author{Jun Ye}
\affiliation{JILA, National Institute of Standards and Technology
and the University of Colorado \\ Department of Physics, University
of Colorado, Boulder, Colorado 80309-0440, USA}
\date{\today}

\begin{abstract}
Cold, neutral hydroxyl radicals are Stark decelerated and confined
within a magnetic trap consisting of two permanent ring magnets. The
OH molecules are trapped in the ro-vibrational ground state at a
density of $\sim$10$^{6}$ cm$^{-3}$ and temperature of 70 mK.
Collisions between the trapped OH sample and supersonic beams of
atomic He and molecular D$_{2}$ are observed and absolute collision
cross sections measured. The He--OH and D$_{2}$--OH center-of-mass
collision energies are tuned from 60 cm$^{-1}$ to 230 cm$^{-1}$ and 145 cm$^{-1}$ to 510 cm$^{-1}$, respectively,
yielding evidence of reduced He--OH inelastic cross sections at energies
below 84 cm$^{-1}$, the OH ground rotational level spacing.
\end{abstract}

\maketitle

Research in the field of cold polar molecules is progressing
rapidly. An array of interesting topics is being developed,
including precision measurement and fundamental
tests~\cite{HindsEDM,Hudson06,LevPRA}, quantum phase
transitions~\cite{Zoller06}, and ultracold
chemistry~\cite{Krems05,HudsonH2CO}. In particular, dipolar
molecules with well defined quantum states will enable us to
exquisitely control their interactions via applied electric fields
\cite{Lewenstein00,BohnFieldLinked}. The long-range, anisotropic
interactions between dipolar molecules lead to new types of
collision dynamics that could be used for novel collective
effects~\cite{Lewenstein02,You04}, quantum state engineering, and
information processing.

Ultracold heteronuclear molecules have been produced via magnetic
association or photoassociation of dual-species ultracold atom
pairs~\cite{JulienneFeshbach,JulienneReview}. Most of these
molecules are in excited rovibrational states, permitting ultracold
atom--molecule collision studies to probe molecular decay processes
\cite{Zirbel08,Hudson08}. Coherent control techniques via exquisite
manipulation of light fields are being developed to drive these
internally excited molecules produced in the initial association
step into deeply bound states with large dipole moments
\cite{Demille05,Ospelkaus08}. High efficiencies expected in the
optical transfer process will lead to unprecedented densities of
ultracold polar molecules for studies of ultracold
dipolar collisions and reactions.

Cold polar molecules in the ground state can be directly produced
via cryogenic buffer gas cooling or Stark deceleration, but at
temperatures typically in the range of 10 mK - 1 K. For example,
buffer gas cooling methods have allowed for magnetic trapping of
NH~\cite{Doyle07} and CaH~\cite{Doyle98} at $\sim1$ K. The presence
of the cooling He atoms naturally led to the studies of He -
molecule collisions~\cite{Doyle08a} and the observation of the
quadratic dependence of the inelastic spin relaxation collision rate
on rotational constant for $^3\Sigma$ NH molecules. Stark
deceleration of supersonic molecular beams readily produces
state-selected, $\sim100$ mK samples at densities of
10$^{6}$--10$^{8}$ cm$^{-3}$~\cite{Meijer99,Bochinski03}. Stark
decelerated molecules have been used in crossed-beam collisions with
atomic species such as Xe~\cite{MeijerOHXe}.

\begin{figure}[t]
\resizebox{0.8\columnwidth}{!}{\includegraphics{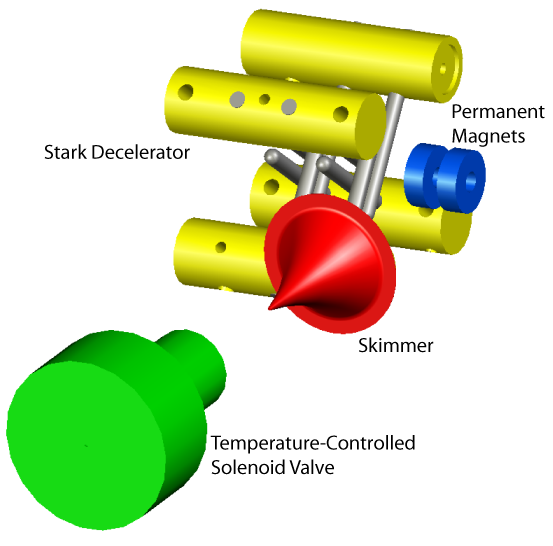}}
\caption{\label{fig1}(color online) Illustration of the magnetic
trap and pulsed valve assemblies. The foreground contains the
temperature-controlled solenoid valve and 1 mm diameter skimmer. The
final stages of the Stark decelerator and permanent ring magnets of
the trap are shown in the background.}
\end{figure}

In this Letter, we describe the confinement of cold dipolar
molecules in a permanent magnetic trap loaded from a Stark
decelerator. The open trap geometry facilitates low-energy
molecule--molecule collision studies. We present total collision
cross sections for D$_{2}$--OH and He--OH collisions within a magnet
trap confining Stark-decelerated OH molecules. The atomic and molecular beams are produced in a supersonic nozzle cooled by liquid
nitrogen, and the He--OH and D$_2$--OH center-of-mass collision energies can be tuned from
60 cm$^{-1}$ to 230 cm$^{-1}$ and 145 cm$^{-1}$ to 510 cm$^{-1}$, respectively. Our magnetic trap design permits the
application of electric dipole fields of tunable
strength~\cite{SawyerMET07} that can be used to study novel dipolar
collisions such as between fully polarized OH and NH$_3$. Both
buffer-gas cooling~\cite{Patterson07} and Stark deceleration produce
a large class of cold polar molecule beams. In combination with the
trap described here, a wide variety of chemically interesting
inter-species collisions can be studied at hitherto unexplored
collision energies.

Importantly, collisions between hydrogen (H, H$_{2}$) and larger
polar molecules (OH, H$_{2}$O, CO, H$_{2}$CO, SiO) are of
astrophysical interest due to their possible role as pumps for
interstellar masers~\cite{ElitzurReview82}. In particular, specific
emission lines of interstellar OH (1720 MHz) and H$_{2}$CO (4.8 GHz)
masers have been attributed to collision-induced inversion by
H$_{2}$~\cite{ElitzurMaser78,H2COMaser03}. By directly comparing
D$_{2}$--OH cross sections with those of the more theoretically
tractable He--OH, we will aid molecular collision theory at energy
scales below 200 cm$^{-1}$.

A pulsed supersonic beam of OH radicals is produced by striking an
electric discharge through a mixture of 27 mbar H$_{2}$O and 1.5 bar
Kr. The resulting OH beam consists of rotationally cold,
$^{2}\Pi_{3/2}$ molecules whose center longitudinal velocity is 490
m/s. The packet passes through a 3 mm diameter skimmer and is then
coupled via an electrostatic hexapole into the 142-stage Stark
decelerator. The Stark decelerator slows weak-field seeking OH
molecules residing in the $|J=3/2,m_{J}=\pm3/2,f\rangle$ state,
where $J$ represents total angular momentum and $m_{J}$ is the
projection of $J$ along the electric ($E$) field axis. The third
quantum number denotes the parity of the state. The design and
operating principle of this decelerator is discussed in previous
work~\cite{Bochinski03,Bochinski04}. The Stark decelerator is
operated at a phase angle $\phi_{0}$ of $50.352^{\circ}$ in order to
slow a 120 mK portion of the OH packet to 36 m/s. When slowing to
velocities below 50 m/s, we observe maximum decelerator efficiency
for $45^{\circ}\leq\phi_{0}\leq55^{\circ}$. Operation at these
intermediate phase angles increases transverse packet confinement
and reduces the coupling between transverse and longitudinal motion
~\cite{Meerakker06,SawyerEPJD}.

\begin{figure}
\includegraphics[height=3.5cm,width=8cm]{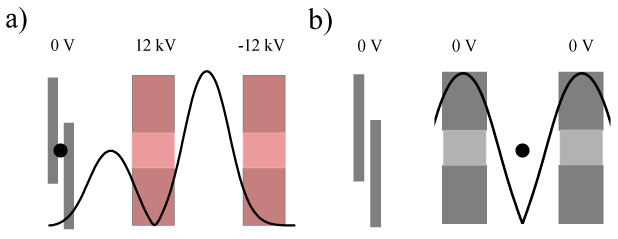}
\caption{\label{seq}(color online) Illustration of the trap loading
sequence. (a) High voltage is applied to the surfaces of the two
permanent ring magnets 1 $\mu$s after the final deceleration stage
(shown at left) is grounded. The OH packet is stopped directly
between the two magnets by the electric field gradient in 400
$\mu$s. The stopping potential due to the applied electric field is
depicted. (b) The magnet surfaces are grounded, leaving the packet
trapped within the displayed permanent magnetic quadrupole
potential.}
\end{figure}

We stop and confine the decelerated 36 m/s OH packet within a
permanent magnetic trap whose center lies 1 cm from the final
decelerator rod pair. This trap, depicted in Fig.~\ref{fig1},
represents a marked improvement over our previous
magneto-electrostatic trap in both design simplicity and ultimate
trapping efficiency~\cite{SawyerMET07}. The magnetic trap is
constructed by mounting two Ni-coated NdFeB permanent ring magnets
in an opposing orientation such that a magnetic quadrupole field is
produced between them. The N42SH rating of these magnets ensures an
operating temperature of up to 150$^{\circ}$C and a residual
magnetization of 1.24 T. The inner and outer radii of the magnets
measure 2 mm and 6 mm, respectively, while their thickness is 4 mm.
The center-to-center magnet spacing of 7 mm in this magnetic trap
yields a longitudinal magnetic ($B$) field gradient of 2 T/cm. This
longitudinal separation matches the extent of the molecular packet
entering the trap region from the Stark decelerator and therefore
maximizes the trap density.

Figure~\ref{seq} illustrates the trap loading sequence used with the
magnetic trap of Fig.~\ref{fig1}. In Fig.~\ref{seq}(a), the Ni
coatings of the ring magnets are charged to $\pm12$ kV precisely 1
$\mu$s after the final deceleration stage is grounded. At this
point, the magnets become high-voltage electrodes and serve as a
final Stark-slowing stage for the 36 m/s molecules. In addition to
the large stopping potential between the magnets, there exists a
smaller potential between the final decelerator rod pair and the
first trap magnet. This barrier reflects the small number of
molecules with longitudinal velocity less than 25 m/s. However, the
barrier's transverse $E$-field gradient serves to confine the slow
molecules as they enter the trap region. The OH packet is brought to
rest directly between the magnets in 400 $\mu$s, at which point the
high voltage is switched off. Those hydroxyl radicals occupying the
weak-magnetic-field-seeking $|3/2,3/2,f\rangle$ state (50\% of the
stopped molecules) are then confined within a magnetic quadrupole
trap measuring $k_{B}\times480$ mK deep in the longitudinal
dimension, where $k_{B}$ is Boltzmann's constant. The trap potential
is shown in Fig.~\ref{seq}(b). We note that a permanent magnet was
used to reflect a molecular beam~\cite{MetsalaPermanent}.

\begin{figure}
\centering
\includegraphics[totalheight=0.45\textheight]{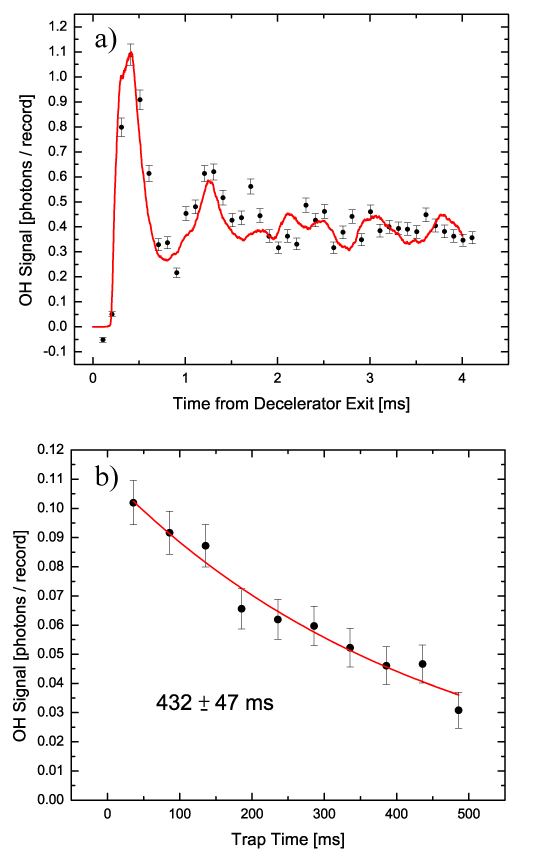}
\caption{\label{TOF}(color online) (a) Time-of-flight data (circles
with error bars) and three-dimensional Monte Carlo simulations
(solid line) corresponding to OH trap loading. Stopping $E$-fields
are switched off at 400 $\mu$s, which leaves 50$\%$ of the stopped
OH molecules trapped in the permanent magnetic quadrupole. (b)
Measurement of the lifetime of OH trapped within the magnetic trap
at a background pressure of $7.5\times10^{-9}$ Torr. A
single-exponential fit (solid line) of the data yields 432 $\pm$ 47
ms.}
\end{figure}

Typical time-of-flight data and corresponding three-dimensional
Monte Carlo simulation results are displayed in Fig.~\ref{TOF}(a).
Decelerated and trapped molecules are detected via laser-induced
fluorescence (LIF). Lenses mounted in-vacuum allow for a
fluorescence collection solid angle of $\sim$0.1 sr. In
Fig.~\ref{TOF}(a), the large peak at 400 $\mu$s is the stopped OH
packet imaged at trap center. Transverse oscillation of the trapped
packet is observed in both data and simulation over 2 ms. The number
and density of trapped OH are measured to be $>$10$^3$ and
$\sim10^{6}$ cm$^{-3}$, respectively. A temperature of 70 mK is
estimated from Monte Carlo simulation, also consistent with the
molecular packet delivered by the Stark decelerator. Due to the
large quadrupole $B$-field present in the trap, only a fraction of
the OH sample is detected as the longitudinal 5 GHz Zeeman shift
near each magnet is larger than our LIF laser linewidth. This effect
is included in the trap density estimates. Figure~\ref{TOF}(b)
displays the observed trap lifetime of 432 $\pm$ 47 ms, limited by
collisions with background gas. The trap chamber pressure of
$7.5\times10^{-9}$ Torr consists of equal parts H$_{2}$O and Kr. We
note that this trap would be ideal for proposed multiple-loading
schemes for molecules such as NH~\cite{MeijerNH01}.

\begin{figure}
\centering
\includegraphics[totalheight=0.45\textheight]{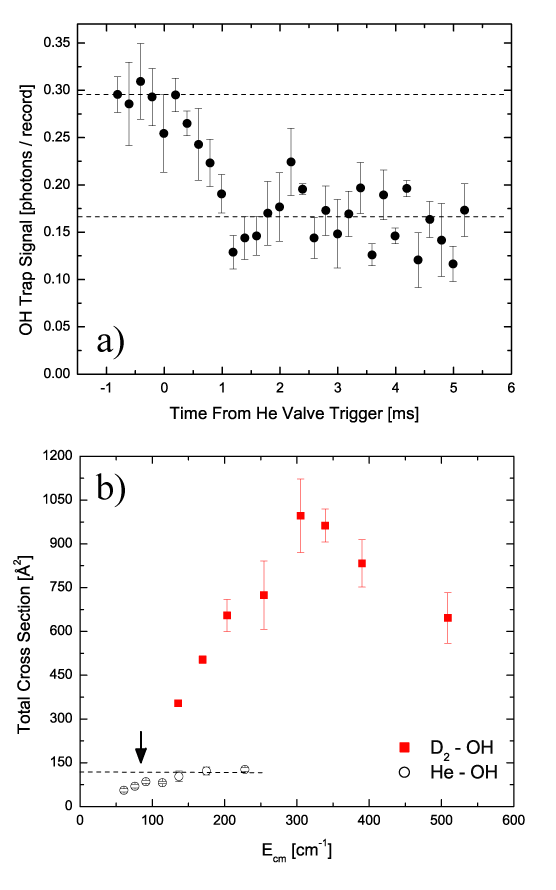}
\caption{\label{scatter}(color online) (a) Time dependence of OH
trap loss due to collisions with a supersonic He beam. Trap density
drops sharply over 1 ms upon He beam collisions, then remains
constant over the time scale shown. The valve is triggered 20 ms
after the magnetic trap is loaded. (b) Total collision cross
sections for He--OH (open circles) and D$_{2}$--OH (squares) as a
function of E$_{cm}$. The decrease in the He--OH cross section at
low energy is attributed to reduced inelastic loss as E$_{cm}$ drops
below the 84 cm$^{-1}$ splitting between the $J=3/2$ and $J=5/2$
states of the OH molecule. The vertical arrow marks the 84 cm$^{-1}$ point while the dashed line highlights this threshold behavior.}
\end{figure}

The open structure of the magnetic trap allows for low
center-of-mass energy (E$_{cm}$) collision studies between the
trapped OH and external molecular or atomic beams. As large electric
fields are used only for initial trap loading, there is no risk of
voltage breakdown when pulsing such beams through the trap.
Furthermore, a relatively small polarizing $E$-field (few kV/cm) can
be applied to the magnets after trap loading without loss of
confined molecules---thereby enabling investigation of dipolar
collisions. Figure~\ref{scatter} displays results from separately
scattering beams of He and D$_{2}$ with the trapped OH sample. For
this work, we place a pulsed solenoid valve (General Valve Series
99) and 1 mm diameter skimmer assembly, as shown in Fig.~\ref{fig1},
such that the skimmed atomic or molecular beam passes directly
between the magnets. The solenoid valve rests in a bath of liquid
nitrogen while its 1 mm output nozzle is heated via 25 turns of
manganin wire to allow for tuning of the beam velocity. All
scattering data is taken at He and D$_{2}$ backing pressures of 2.0
and 2.7 bar, respectively. As expected, we observe that the beam
velocities scale as the square-root of the nozzle temperature. The
E$_{cm}$ of the He--OH and D$_{2}$--OH systems can therefore be
tuned to minima of $\sim$60 cm$^{-1}$ and $\sim$145 cm$^{-1}$, respectively. Measurements of beam flux
and velocity are made using a fast ionization gauge and a
microphone-based pressure sensor placed 13 cm apart. The uncertainty
in the inter-species calibration of the ionization gauge is
$\sim$10$\%$. The measured 8$\%$ velocity spread of the supersonic
He beam gives a collision energy resolution of 9 cm$^{-1}$ at the
lowest nozzle temperature.

Figure~\ref{scatter}(a) displays the time dependence of OH trap loss
as a supersonic beam of He traverses the magnetic trap. Trap density
drops sharply over the first $\sim$1 ms after the solenoid valve is
fired, then remains constant over the time scale shown. Although the
partial pressure of the scattering gas in the trap chamber rises as
the supersonic beam scatters from the chamber walls, we measure an
OH trap lifetime of $>$50 ms following the initial collision. Such
vastly different time scales allow us to differentiate between trap
loss due to the supersonic beam and that resulting from background
gas collisions. Trap loss at a given nozzle temperature is measured
by repeatedly comparing OH population 1 ms \textit{before} and 2 ms
\textit{after} the solenoid valve is triggered. The total cross
sections of Fig.~\ref{scatter}(b) are determined from the trap loss
data by normalization to the corresponding beam flux measured with
the fast ionization gauge at a given nozzle temperature. We also
find that the trap loss scales linearly with the beam flux,
confirming that we are operating in the single-collision regime.

An important advantage inherent to using trapped molecules as the
scattering target is the ability to directly measure the absolute
collision cross section. To do this, we use a leak valve to fill the
trap chamber with a known pressure of He gas, then directly measure
the OH trap lifetime at that pressure. The temperature of the
chamber walls is 298 K, placing the E$_{cm}$ of the thermal He--OH
collisions at $\sim$250 cm$^{-1}$. The data point for He--OH
collisions at 230 cm$^{-1}$ from Fig.~\ref{scatter}(b) is then
scaled to this absolute cross section measuring 127 $\pm$ 18
\AA$^{2}$.

For OH molecules in their ground electronic and vibrational state,
the energy splitting of the two lowest lying rotational levels,
$J=3/2$ and $J=5/2$, is 84 cm$^{-1}$. In a crossed-beam experiment,
an abrupt decrease in the Xe--OH inelastic cross section was
observed as E$_{cm}$ was tuned below this value~\cite{MeijerOHXe}.
The He data of Fig.~\ref{scatter}(b) possesses this feature. Because
the magnetic trapping potential is sensitive to the internal state
of the OH molecule, we cannot differentiate between trap loss due to
elastic or inelastic collisions. Nevertheless, such a sudden
decrease of the total cross section below 84 cm$^{-1}$ is indicative
of threshold behavior. The collision cross section of D$_2$--OH is
larger than that of He--OH. This is understandable since the D$_2$
beam is an ortho/para mixture and contains a large fraction of
frozen-in $J=1$ population. The ratio of the two cross sections at
230 cm$^{-1}$ is consistent with previous H$_{2}$--OH and He--OH
pressure-broadening measurements made at a temperature of 298
K~\cite{OHPressureBroadening}. The enhanced collision cross sections
for D$_2$--OH result from the quadrupole moment of the D$_2$ $J=1$
state that can interact strongly at long range with the OH dipole.
Yet another striking feature of the collision data is the pronounced
peak in the D$_{2}$--OH cross section for E$_{cm}\sim305$ cm$^{-1}$.
Although more theoretical consideration is warranted, the 300
cm$^{-1}$ $J=3$ $\leftarrow$ $J=1$ transition of the
$^{1}\Sigma_{g}^{+}$ D$_{2}$ molecule may be contributing to the
inelastic cross section at this energy. Another possible explanation
is collision-induced decay of $J=2$ molecules present in an
imperfect D$_{2}$ supersonic expansion.

In conclusion, we demonstrate a new permanent magnetic trap design
that confines a dense sample of cold OH molecules as a cold
collision target. Velocity-tunable supersonic beams of He and
D$_{2}$ intersecting the magnetically trapped cold OH molecules
yield absolute collision cross sections over an energy range of 60
cm$^{-1}$ to 230 cm$^{-1}$ and 145 cm$^{-1}$ to 510 cm$^{-1}$,
respectively. Threshold behavior is observed in He--OH collisions,
and an enhancement of inelastic cross sections is seen in the
D$_{2}$--OH system near 305 cm$^{-1}$. Having demonstrated the
usefulness of this permanent magnetic trap for collision studies,
future goals include using colder continuous beams of polar
molecules as the colliding partner for OH. With the ability to apply
a sufficiently strong polarizing field within the magnetic trap
\cite{Krems06}, we aim to reach sufficiently low E$_{cm}$ to observe
dipole-dipole interactions.

We acknowledge DOE, NSF, and NIST for funding support. We thank E.
Meyer, J. L. Bohn, B. Lev, and J. M. Hutson for stimulating
discussions and T. Keep and H. Green for technical assistance.


\begin{thebibliography}{33}
\expandafter\ifx\csname
natexlab\endcsname\relax\def\natexlab#1{#1}\fi
\expandafter\ifx\csname bibnamefont\endcsname\relax
  \def\bibnamefont#1{#1}\fi
\expandafter\ifx\csname bibfnamefont\endcsname\relax
  \def\bibfnamefont#1{#1}\fi
\expandafter\ifx\csname citenamefont\endcsname\relax
  \def\citenamefont#1{#1}\fi
\expandafter\ifx\csname url\endcsname\relax
  \def\url#1{\texttt{#1}}\fi
\expandafter\ifx\csname urlprefix\endcsname\relax\def\urlprefix{URL
}\fi \providecommand{\bibinfo}[2]{#2}
\providecommand{\eprint}[2][]{\url{#2}}

\bibitem[{\citenamefont{Hudson et~al.}(2002)}]{HindsEDM}
\bibinfo{author}{\bibfnamefont{J.~J.} \bibnamefont{Hudson}}
  \bibnamefont{et~al.}, \bibinfo{journal}{Phys. Rev. Lett.}
  \textbf{\bibinfo{volume}{89}}, \bibinfo{pages}{023003}
  (\bibinfo{year}{2002}).

\bibitem[{\citenamefont{Hudson et~al.}(2006{\natexlab{a}})}]{Hudson06}
\bibinfo{author}{\bibfnamefont{E.~R.} \bibnamefont{Hudson}}
  \bibnamefont{et~al.}, \bibinfo{journal}{Phys. Rev. Lett.}
  \textbf{\bibinfo{volume}{96}}, \bibinfo{pages}{143004}
  (\bibinfo{year}{2006}{\natexlab{a}}).

\bibitem[{\citenamefont{Lev et~al.}(2006)}]{LevPRA}
\bibinfo{author}{\bibfnamefont{B.~L.} \bibnamefont{Lev}} \bibnamefont{et~al.},
  \bibinfo{journal}{Phys. Rev. A} \textbf{\bibinfo{volume}{74}},
  \bibinfo{pages}{061402(R)} (\bibinfo{year}{2006}).

\bibitem[{\citenamefont{Micheli et~al.}(2006)}]{Zoller06}
\bibinfo{author}{\bibfnamefont{A.}~\bibnamefont{Micheli}} \bibnamefont{et~al.},
  \bibinfo{journal}{Nature Physics} \textbf{\bibinfo{volume}{2}},
  \bibinfo{pages}{341} (\bibinfo{year}{2006}).

\bibitem[{\citenamefont{Krems}(2005)}]{Krems05}
\bibinfo{author}{\bibfnamefont{R.~V.} \bibnamefont{Krems}},
  \bibinfo{journal}{Intern. Rev. Phys. Chem.} \textbf{\bibinfo{volume}{24}},
  \bibinfo{pages}{99} (\bibinfo{year}{2005}).

\bibitem[{\citenamefont{Hudson et~al.}(2006{\natexlab{b}})}]{HudsonH2CO}
\bibinfo{author}{\bibfnamefont{E.~R.} \bibnamefont{Hudson}}
  \bibnamefont{et~al.}, \bibinfo{journal}{Phys. Rev. A}
  \textbf{\bibinfo{volume}{76}}, \bibinfo{pages}{063404}
  (\bibinfo{year}{2006}{\natexlab{b}}).

\bibitem[{\citenamefont{Santos et~al.}(2000)}]{Lewenstein00}
\bibinfo{author}{\bibfnamefont{L.}~\bibnamefont{Santos}} \bibnamefont{et~al.},
  \bibinfo{journal}{Phys. Rev. Lett.} \textbf{\bibinfo{volume}{85}},
  \bibinfo{pages}{1791} (\bibinfo{year}{2000}).

\bibitem[{\citenamefont{Avdeenkov et~al.}(2004)\citenamefont{Avdeenkov,
  Bortolotti, and Bohn}}]{BohnFieldLinked}
\bibinfo{author}{\bibfnamefont{A.~V.} \bibnamefont{Avdeenkov}},
  \bibinfo{author}{\bibfnamefont{D.~C.~E.} \bibnamefont{Bortolotti}},
  \bibnamefont{and} \bibinfo{author}{\bibfnamefont{J.~L.} \bibnamefont{Bohn}},
  \bibinfo{journal}{Phys. Rev. A} \textbf{\bibinfo{volume}{69}},
  \bibinfo{pages}{012710} (\bibinfo{year}{2004}).

\bibitem[{\citenamefont{Goral et~al.}(2002)\citenamefont{Goral, Santos, and
  Lewenstein}}]{Lewenstein02}
\bibinfo{author}{\bibfnamefont{K.}~\bibnamefont{Goral}},
  \bibinfo{author}{\bibfnamefont{L.}~\bibnamefont{Santos}}, \bibnamefont{and}
  \bibinfo{author}{\bibfnamefont{M.}~\bibnamefont{Lewenstein}},
  \bibinfo{journal}{Phys. Rev. Lett.} \textbf{\bibinfo{volume}{88}},
  \bibinfo{pages}{170406} (\bibinfo{year}{2002}).

\bibitem[{\citenamefont{Yi et~al.}(2004)\citenamefont{Yi, You, and Pu}}]{You04}
\bibinfo{author}{\bibfnamefont{S.}~\bibnamefont{Yi}},
  \bibinfo{author}{\bibfnamefont{L.}~\bibnamefont{You}}, \bibnamefont{and}
  \bibinfo{author}{\bibfnamefont{H.}~\bibnamefont{Pu}}, \bibinfo{journal}{Phys.
  Rev. Lett.} \textbf{\bibinfo{volume}{93}}, \bibinfo{pages}{040403}
  (\bibinfo{year}{2004}).

\bibitem[{\citenamefont{Kohler et~al.}(2006)}]{JulienneFeshbach}
\bibinfo{author}{\bibfnamefont{T.}~\bibnamefont{Kohler}} \bibnamefont{et~al.},
  \bibinfo{journal}{Rev. Mod. Phys.} \textbf{\bibinfo{volume}{78}},
  \bibinfo{pages}{1311} (\bibinfo{year}{2006}).

\bibitem[{\citenamefont{Jones et~al.}(2006)}]{JulienneReview}
\bibinfo{author}{\bibfnamefont{K.~M.} \bibnamefont{Jones}}
  \bibnamefont{et~al.}, \bibinfo{journal}{Rev. Mod. Phys.}
  \textbf{\bibinfo{volume}{78}}, \bibinfo{pages}{483} (\bibinfo{year}{2006}).

\bibitem[{\citenamefont{Zirbel et~al.}(2008)}]{Zirbel08}
\bibinfo{author}{\bibfnamefont{J.~J.} \bibnamefont{Zirbel}}
  \bibnamefont{et~al.}, \bibinfo{journal}{Phys. Rev. Lett.}
  \textbf{\bibinfo{volume}{100}}, \bibinfo{pages}{143201}
  (\bibinfo{year}{2008}).

\bibitem[{\citenamefont{Hudson et~al.}(2008)}]{Hudson08}
\bibinfo{author}{\bibfnamefont{E.~R.} \bibnamefont{Hudson}}
  \bibnamefont{et~al.}, \bibinfo{journal}{Phys. Rev. Lett.}
  \textbf{\bibinfo{volume}{100}}, \bibinfo{pages}{203201}
  (\bibinfo{year}{2008}).

\bibitem[{\citenamefont{Sage et~al.}(2005)}]{Demille05}
\bibinfo{author}{\bibfnamefont{J.}~\bibnamefont{Sage}} \bibnamefont{et~al.},
  \bibinfo{journal}{Phys. Rev. Lett.} \textbf{\bibinfo{volume}{94}},
  \bibinfo{pages}{203001} (\bibinfo{year}{2005}).

\bibitem[{\citenamefont{Ospelkaus et~al.}(2008)}]{Ospelkaus08}
\bibinfo{author}{\bibfnamefont{S.}~\bibnamefont{Ospelkaus}}
  \bibnamefont{et~al.} (\bibinfo{year}{2008}), \eprint{arXiv/0802.1093; Nature
  Physics, in press}.

\bibitem[{\citenamefont{Campbell et~al.}(2007)}]{Doyle07}
\bibinfo{author}{\bibfnamefont{W.~C.} \bibnamefont{Campbell}}
  \bibnamefont{et~al.}, \bibinfo{journal}{Phys. Rev. Lett.}
  \textbf{\bibinfo{volume}{98}}, \bibinfo{pages}{213001}
  (\bibinfo{year}{2007}).

\bibitem[{\citenamefont{Weinstein et~al.}(1998)}]{Doyle98}
\bibinfo{author}{\bibfnamefont{J.}~\bibnamefont{Weinstein}}
  \bibnamefont{et~al.}, \bibinfo{journal}{Nature}
  \textbf{\bibinfo{volume}{395}}, \bibinfo{pages}{148} (\bibinfo{year}{1998}).

\bibitem[{\citenamefont{Campbell et~al.}(2008)}]{Doyle08a}
\bibinfo{author}{\bibfnamefont{W.~C.} \bibnamefont{Campbell}}
  \bibnamefont{et~al.} (\bibinfo{year}{2008}), \eprint{arXiv:0804.0265v1}.

\bibitem[{\citenamefont{{H.L. Bethlem} et~al.}(1999)\citenamefont{{H.L.
  Bethlem}, Berden, and Meijer}}]{Meijer99}
\bibinfo{author}{\bibnamefont{{H.L. Bethlem}}},
  \bibinfo{author}{\bibfnamefont{G.}~\bibnamefont{Berden}}, \bibnamefont{and}
  \bibinfo{author}{\bibfnamefont{G.}~\bibnamefont{Meijer}},
  \bibinfo{journal}{Phys. Rev. Lett.} \textbf{\bibinfo{volume}{83}},
  \bibinfo{pages}{1558} (\bibinfo{year}{1999}).

\bibitem[{\citenamefont{Bochinski et~al.}(2003)}]{Bochinski03}
\bibinfo{author}{\bibfnamefont{J.}~\bibnamefont{Bochinski}}
  \bibnamefont{et~al.}, \bibinfo{journal}{Phys. Rev. Lett.}
  \textbf{\bibinfo{volume}{91}}, \bibinfo{pages}{243001}
  (\bibinfo{year}{2003}).

\bibitem[{\citenamefont{Gilijamse et~al.}(2006)}]{MeijerOHXe}
\bibinfo{author}{\bibfnamefont{J.~J.} \bibnamefont{Gilijamse}}
  \bibnamefont{et~al.}, \bibinfo{journal}{Science}
  \textbf{\bibinfo{volume}{313}}, \bibinfo{pages}{1617} (\bibinfo{year}{2006}).

\bibitem[{\citenamefont{Sawyer et~al.}(2007)}]{SawyerMET07}
\bibinfo{author}{\bibfnamefont{B.~C.} \bibnamefont{Sawyer}}
  \bibnamefont{et~al.}, \bibinfo{journal}{Phys. Rev. Lett.}
  \textbf{\bibinfo{volume}{98}}, \bibinfo{pages}{253002}
  (\bibinfo{year}{2007}).

\bibitem[{\citenamefont{Patterson and Doyle}(2007)}]{Patterson07}
\bibinfo{author}{\bibfnamefont{D.}~\bibnamefont{Patterson}} \bibnamefont{and}
  \bibinfo{author}{\bibfnamefont{J.~M.} \bibnamefont{Doyle}},
  \bibinfo{journal}{J. Chem. Phys.} \textbf{\bibinfo{volume}{126}},
  \bibinfo{pages}{154307} (\bibinfo{year}{2007}).

\bibitem[{\citenamefont{Elitzur}(1982)}]{ElitzurReview82}
\bibinfo{author}{\bibfnamefont{M.}~\bibnamefont{Elitzur}},
  \bibinfo{journal}{Rev. Mod. Phys.} \textbf{\bibinfo{volume}{54}},
  \bibinfo{pages}{1225} (\bibinfo{year}{1982}).

\bibitem[{\citenamefont{Guibert et~al.}(1978)}]{ElitzurMaser78}
\bibinfo{author}{\bibfnamefont{J.}~\bibnamefont{Guibert}} \bibnamefont{et~al.},
  \bibinfo{journal}{Astron. Astrophys.} \textbf{\bibinfo{volume}{62}},
  \bibinfo{pages}{305} (\bibinfo{year}{1978}).

\bibitem[{\citenamefont{Hoffman et~al.}(2003)}]{H2COMaser03}
\bibinfo{author}{\bibfnamefont{I.~M.} \bibnamefont{Hoffman}}
  \bibnamefont{et~al.}, \bibinfo{journal}{Astrophys. J.}
  \textbf{\bibinfo{volume}{598}}, \bibinfo{pages}{1061} (\bibinfo{year}{2003}).

\bibitem[{\citenamefont{Bochinksi et~al.}(2004)}]{Bochinski04}
\bibinfo{author}{\bibfnamefont{J.}~\bibnamefont{Bochinski}}
  \bibnamefont{et~al.}, \bibinfo{journal}{Phys. Rev. A}
  \textbf{\bibinfo{volume}{70}}, \bibinfo{pages}{043410}
  (\bibinfo{year}{2004}).

\bibitem[{\citenamefont{van~de Meerakker et~al.}(2006)}]{Meerakker06}
\bibinfo{author}{\bibfnamefont{S.~Y.~T.} \bibnamefont{van~de Meerakker}}
  \bibnamefont{et~al.}, \bibinfo{journal}{Phys. Rev. A}
  \textbf{\bibinfo{volume}{73}}, \bibinfo{pages}{023401}
  (\bibinfo{year}{2006}).

\bibitem[{\citenamefont{Sawyer et~al.}()}]{SawyerEPJD}
\bibinfo{author}{\bibfnamefont{B.~C.} \bibnamefont{Sawyer}}
  \bibnamefont{et~al.}, \bibinfo{journal}{Eur. Phys. J. D} \textbf{\bibinfo{volume}{48}},
  \bibinfo{pages}{197} (\bibinfo{year}{2008}).

\bibitem[{\citenamefont{van~de Meerakker et~al.}(2001)}]{MeijerNH01}
\bibinfo{author}{\bibfnamefont{S.~Y.~T.} \bibnamefont{van~de Meerakker}}
  \bibnamefont{et~al.}, \bibinfo{journal}{Phys. Rev. A}
  \textbf{\bibinfo{volume}{64}}, \bibinfo{pages}{041401(R)}
  (\bibinfo{year}{2001}).

\bibitem[{\citenamefont{Metsala et~al.}()}]{MetsalaPermanent}
\bibinfo{author}{\bibfnamefont{M.} \bibnamefont{Metsala}}
  \bibnamefont{et~al.}, \bibinfo{note}{arXiv:0802.2902} (\bibinfo{year}{2008}).

\bibitem[{\citenamefont{Park et~al.}(1995)}]{OHPressureBroadening}
\bibinfo{author}{\bibfnamefont{K.}~\bibnamefont{Park}} \bibnamefont{et~al.},
  \bibinfo{journal}{J. Quant. Spectrosc. Radiat. Transfer}
  \textbf{\bibinfo{volume}{55}}, \bibinfo{pages}{285} (\bibinfo{year}{1995}).

\bibitem[{\citenamefont{Krems}(2006)}]{Krems06}
\bibinfo{author}{\bibfnamefont{R.~V.} \bibnamefont{Krems}},
  \bibinfo{journal}{Phys. Rev. Lett.} \textbf{\bibinfo{volume}{96}},
  \bibinfo{pages}{123202} (\bibinfo{year}{2006}).

\end{thebibliography}
\end{document}